# Multi-Swing Transient Stability of Synchronous Generators and IBR Combined Generation Systems

Songhao Yang, *Member*, *IEEE*, Bingfang Li, *Student Member*, *IEEE*, Zhiguo Hao, *Senior Member*, *IEEE*, Yiwen Hu, Huan Xie, Tianqi Zhao, and Baohui Zhang, *Fellow, IEEE*

*Abstract*—In traditional views, the build-up of accelerating energy during faults can cause the well-known first-swing angle instability in synchronous generators (SGs). Interestingly, this letter presents a new insight that the accumulation of decelerating energy due to the low voltage ride-through (LVRT) and recovery control of grid-following inverter-based resources (GFL-IBRs), might also result in transient angle instability in SGs. The transient energy accumulated during angle-decreasing swing transforms into the acceleration energy of the subsequent swing, hence such phenomena often manifest as multi-swing instability. Both theoretical analysis and simulation support these findings.

## I. INTRODUCTION

Combining GFL-IBRs and SGs for centralized, long-distance transmission enhances resource flexibility and network strength. However, IBRs exhibit complex effects on the transient stability of such combined systems. Ref. [1] introduces a transient stability criterion to identify operational areas impacting system stability, while ref. [2] notes that IBRs' gradual active current restoration post-LVRT can enhance the first-swing stability of nearby SGs. Additionally, ref. [3] explores the transient interaction of the GFL-IBRs and virtual synchronous generators (VSG) combined systems.

The existing studies mainly focus on the first-swing stability. However, IBR's time-varying output during LVRT and recovery control can change the unbalanced power of nearby SGs, thus affecting their multi-swing dynamics. Inspired by such an idea, this letter explores the multi-swing transient instability of SGs affected by adjacent GFL-IBRs. Utilizing the transient energy function method, this instability phenomenon is explained by the 'Spring Effect', and these findings are verified by simulations.

## II. SYSTEM MODELING

Fig. 1 showcases a typical topology of the IBR-SG combined system, and Fig. 2 presents the rotating vectors of the coordinate system. In Fig. 2, the $d_s$-$q_s$ and $x$-$y$ reference frames are counterclockwise rotating with the angular speed of $\omega_s$ and $\omega_g$ respectively. The $d_s$-axis leads the $x$-axis by an angle $\delta$, representing SG's rotor angle, while the $d$-axis leads by an angle $\theta$, as provided by the PLL. Reference [1], [4] suggest that the dynamics of PLL (tens of Hz) can be ignored on the rotor

This work was supported by Science and Technology Project of State Grid Jibei Electric Power Co., Ltd "Research on the coordinated control technology of grid-forming renewable energy station with new synchronous condenser" (B3018K240005)
S. Yang, B. Li, Z. Hao, Y. Hu, and B. Zhang are with Xi'an Jiaotong University, Xi'an, China (e-mail: {songhaoyang, zhghao, bhzhang}@xjtu.edu.cn, {libingfang, huyiwen}@stu.xjtu.edu.cn). H. Xie and T. Zhao are with State Grid Hebei Electric Power Co.

motion timescale (0.1~2 Hz). Thus, the dynamic equation of the SG can be written as

$$\begin{cases} \dot{\delta} = \omega_g(\omega_s - 1) = \omega_g \Delta\omega \\ T_J \Delta\dot{\omega} = P_M - P_E - D\Delta\omega \end{cases}, \quad (1)$$

where $P_M$, $P_E$, $T_J$, and $D$ are the mechanical power, electrical power, time constant inertia, and damping coefficient of SG, respectively. As per the definition $P_E$=Re($E_s\angle\delta \cdot I_s\angle\varphi_s$),

$$P_E = \underbrace{E_s U_g |Y_{sg}| \sin\delta}_{P_{sg}} - \underbrace{\alpha E_s I_w \cos(\delta-\varphi_w)}_{P_w}, \quad (2)$$

where $\alpha=Y_s/(Y_s+Y_g)$, $Y_{sg}=Y_sY_g/(Y_s+Y_g)$. $P_w$ represents the power coupling term between SG and IBR. Define $I_w^{d_s}=I_w\cos(\delta-\varphi_w)$ as the projection of IBR's output current vector $I_w\angle\varphi_w$ on the $d_s$-axis, which determines the value of $P_w$. Changes in $P_w$ further influence $P_E$, thereby affecting the rotor dynamics of the SG.

In the steady state with unity power factor control, $\varphi_w=\theta$. During transients, if the IBR generates reactive power, $\varphi_w$ adjusts to $\theta-\eta$ as shown in Fig. 2, where $\eta$ is defined as

$$\eta = \arctan\left(|i_w^q / i_w^d|\right). \quad (3)$$

In (3), $i_w^d$ and $i_w^q$ are the active and reactive output current of the IBR.

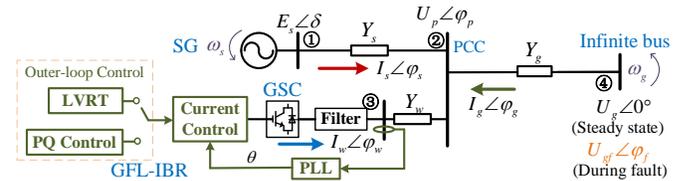

**Fig. 1.** Typical topology of SG-IBR combined generation systems.

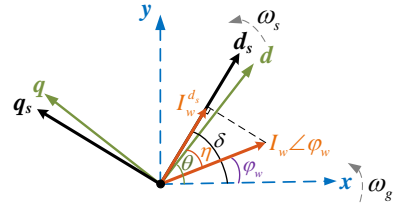

**Fig. 2.** Relationships of different frames in the vector space.

## III. MULTI-SWING TRANSIENT STABILITY ANALYSIS

### A. Multi-Swing Instability Caused by IBR's LVRT Control

The energy function of the system (1) can be formulated as

$$V(\delta, \Delta\omega) = \frac{1}{2}T_J\omega_g\Delta\omega^2 + \int_{\delta_s}^{\delta}(P_{sg} - P_w - P_M)d\delta, \quad (4)$$

where $\delta_s$ is denoted as the stable equilibrium point (SEP) of SG:

$$\delta_s = \arcsin(P_M + P_w)/(E_s U_g |Y_{sg}|). \quad (5)$$

Assuming a grid fault shifts $U_g\angle 0°$ to $U_{gf}\angle\varphi_f$ while triggers the IBR's LVRT control, both components $P_{sg}$ and $P_w$ within $P_E$ would change, labelled as $P_{sgf}$ and $P_{wf}$.

$$P_{sgf} = E_s U_{gf} |Y_{sg}| \sin(\delta + \varphi_f), P_{wf} = \alpha E_s I_{wf}^{d_s}. \quad (6)$$

As shown in Fig. 3, the intersection of curves $P_e(\delta)$ and $P_M+P_w$, i.e. Point A, represents the Stable Equilibrium Point (SEP). Once a fault occurs, $P_{sg}(\delta)$ shifts to $P_{sgf}(\delta)$. Since $\delta$ cannot change abruptly, the operating point moves from Point A to B. After LVRT control activation, due to an increase in $i_w^q$ and usually a decrease in $i_w^d$, the projection of the IBR output current vector on the $d_s$ axis, $I_{wf}^{ds}$, decreases from its steady state value $I_w^{ds}$, resulting in $P_{wf}<P_w$. According to (1)-(2), if $P_{sgf}(\delta)<P_M+P_{wf}$ during the fault, then $\Delta\omega>0$, causing the SG's rotor to accelerate; otherwise, $\Delta\omega<0$, causing it to decelerate.

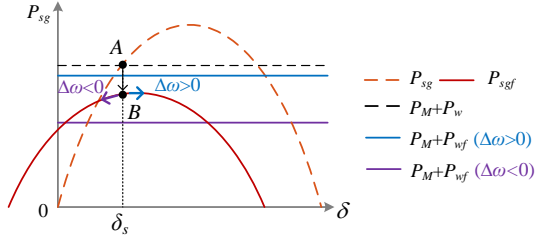

Fig. 3. Power-angle plane of the SG during IBR's LVRT.

Define $\Delta P_{sgf}=P_{sgf}-P_{sg}$ and $\Delta P_{wf}=P_{wf}-P_w$, generally $\Delta P_{wf}<0$. The derivation of $V$ in (4) during fault could be derived as

$$\frac{dV(\delta,\Delta\omega)}{dt} = \frac{\partial V}{\partial \delta}\frac{d\delta}{dt} + \frac{\partial V}{\partial \Delta\omega}\frac{d\Delta\omega}{dt}$$
$$= \omega_g\left(\Delta P_{wf}\Delta\omega - \Delta P_{sgf}\Delta\omega - D\Delta\omega^2\right). \quad (7)$$

Eq.(7) indicates that the effects of $\Delta P_{wf}$ vary with $\Delta\omega$. Specifically, if $\Delta\omega>0$, then $\Delta P_{wf}\Delta\omega<0$, thus $\Delta P_{wf}$ exhibits a positive damping effect, reducing $V$ as a result. In contrast, if $\Delta\omega<0$, the damping effect of $\Delta P_{wf}$ turns negative since $\Delta P_{wf}\Delta\omega>0$, thereby contributing to the increase of $V$.

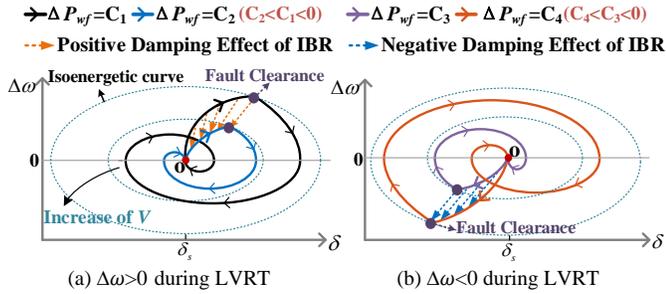

(a) $\Delta\omega>0$ during LVRT   (b) $\Delta\omega<0$ during LVRT
Fig. 4. Phase trajectories of SG during IBR's LVRT control.

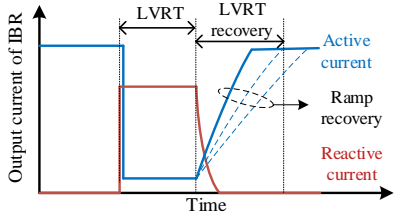

Fig. 5. IBR's output under the LVRT and recovery control.

Assuming a sign change in $\Delta\omega$ marks the end of a rotor swing. Fig. 4 illustrates these swing-dependent damping effects of IBR, where the magnitude of $\Delta P_{wf}$ is characterized by $C_1$ to $C_4$. Fig. 4(a) shows that during faults, if SG accelerates, the accelerating energy is mitigated by the positive damping effect of $\Delta P_{wf}$, reducing the risk of first-swing instability. Conversely, as shown in Fig. 4(b), if the SG decelerates in the first swing, the negative damping effect of $\Delta P_{wf}$ exacerbates the risk of acceleration instability in the subsequent swing by increasing the transient (decelerating) energy during faults.

### B. Multi-Swing Instability Caused by IBR's Recovery Control

After the fault is cleared, $P_{sgf}$ reverts to $P_{sg}$ and IBR's output gradually recovers under the LVRT recovery control (such as ramp recovery control in Fig. 5). During this stage, the power coupling term, marked as $P_{wr}$, satisfies $P_{wf}<P_{wr}<P_w$. Thus, d$V$/d$t$ in (7) is updated to

$$\dot{V}(\delta,\Delta\omega) = \Delta P_{wr}\omega_g\Delta\omega - D\omega_g\Delta\omega^2. \quad (8)$$

Define the change in transient energy over one cycle $T$ as $\Delta V$:

$$\Delta V = \int_{V_0}^{V} dV = \underbrace{\int_0^T (\Delta P_{wr}\omega_g\Delta\omega)dt}_{\Delta V_w} - \underbrace{\int_0^T (D\omega_g\Delta\omega^2)dt}_{\Delta V_D} \quad (9)$$

where $V_0$ denotes the transient energy at the start of this cycle (marked as point A in Fig. 6). By the end of the cycle (marked as point C in Fig. 6), $V$ can be expressed as

$$V = V_0 - V_D + \omega_g\int_{\delta_A}^{\delta_B}\Delta P_w(t)d\delta + \omega_g\int_{\delta_B}^{\delta_C}\Delta P_w(t)d\delta \quad (10)$$

Define $\Delta P_{wr}(t)=P_{wr}(t)-P_w$. It is a time-increasing function and remains negative during the recovery phase. Besides, it's also a monotonic function of $\delta$ respectively in the intervals A→B and B→C. According to the Mean Value Theorem for Integrals,

$$\int_{\delta_A}^{\delta_B}\Delta P_{wr}(t)d\delta = \overline{\Delta P_{wr1}}(\delta_B - \delta_A), \quad (11)$$

$$\int_{\delta_B}^{\delta_C}\Delta P_{wr}(t)d\delta = \overline{\Delta P_{wr2}}(\delta_C - \delta_B), \quad (12)$$

where $\overline{\Delta P_{wr1}}$ and $\overline{\Delta P_{wr2}}$ are both constants and satisfies $\overline{\Delta P_{wr1}} < \overline{\Delta P_{wr2}} \leq 0$. According to (7), similar to the LVRT stage, $\overline{\Delta P_{wr1}}$ exerts a negative damping effect during A→B because $\overline{\Delta P_{wr1}}\Delta\omega>0$, leading $V$ to increase. In contrast, $\overline{\Delta P_{wr2}}\Delta\omega<0$ during B→C, resulting in a reduction of $V$.

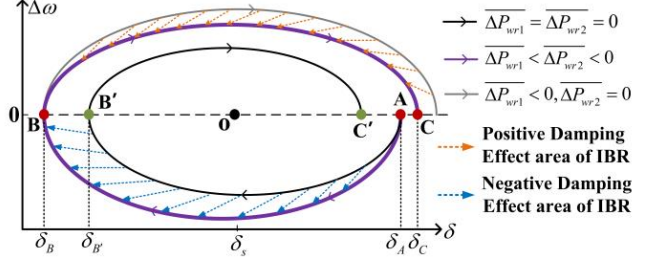

Fig. 6. Phase trajectories of SG during IBR's LVRT recovery.

Substitute (10)-(12) into (9), we can obtain:

$$\Delta V = \underbrace{(\overline{\Delta P_{wr2}} - \overline{\Delta P_{wr1}})(\delta_A - \delta_B) + \overline{\Delta P_{wr2}}(\delta_C - \delta_A)}_{\Delta V_w} - \Delta V_D \quad (13)$$

According to (13), the cyclic energy change $\Delta V$ is determined by the sum of the energy change $\Delta V_w$ from IBR's damping effects and the energy dissipation $\Delta V_D$ from SG's physical damping. If $\delta_A$-$\delta_C<0$, then $\Delta V>0$. Since $\Delta V_D>0$, it follows that $\Delta V_w>0$. If $\delta_A$-$\delta_C>0$, $\Delta V_w>0$ also holds because $\overline{\Delta P_{wr1}} < \overline{\Delta P_{wr2}} \leq 0$ and $\delta_A$-$\delta_B>0$. This indicates that IBR's negative damping effect is predominant when $\Delta\omega<0$, contributing to the increase in $\Delta V$. If $\Delta V_w>\Delta V_D$ within one cycle, the multi-swing stability of SGs will deteriorate, as shown in Fig. 6. Negative $\overline{\Delta P_{wr1}}$ lead to an increase in $V$ during A→B, compared to $\overline{\Delta P_{wr1}}=0$. Despite a decrease in $V$ during B→C, the increase in $V$ is more significant within one cycle. Thus, the time-varying damping effects of the IBR during the recovery process increase transient (decelerating) energy during angle-decreasing swings and potentially lead to instability in the subsequent swing.

*C. Instability Mechanism Analogous to the "Spring Effect"*

The mechanisms behind the first-swing instability induced by excessive mechanical power (Mode 1) and the multi-swing instability caused by LVRT and recovery control of IBRs (Mode 2) are markedly different. This can be analogized to the "Spring Effect" seen in a pull-back toy car, as depicted in Fig. 7. The finish line ($\delta_{uep}$ in Fig. 7(a)) represents SG's unstable equilibrium point (UEP). One way of play involves propelling the toy car directly to cross the finish line, where the forward thrust represents the accelerating energy of SGs accumulated during a large disturbance (corresponding to the scenario in Fig. 4(a)). The other way is to first pull the car back and then release it. When the car is retracted, the internal spring stores potential energy, which then converts to kinetic energy upon release. This backward thrust mirrors the decelerating energy in the angle-decreasing direction of SGs, resulting from the negative damping effect of IBR's LVRT and slow recovery control, as illustrated in Fig. 4(b) and Fig. 6.

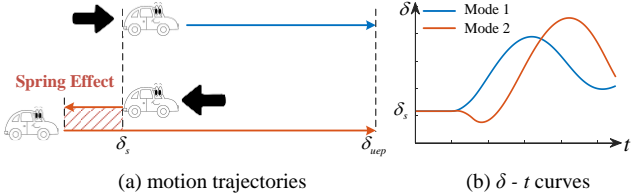

(a) motion trajectories  (b) $\delta$ - $t$ curves
**Fig. 7.** Pull-back toy car analogy for SG's multi-swing instability.

## IV. NUMERICAL RESULTS

A simplified IBR-SG system test model is established in DIgSILENT, with system parameters provided in TABLE I. At $t$=0.5s, a three-phase permanent short circuit occurs on one of the double circuit lines, cleared by protection at $t$=0.7s. To verify the impact of GFL-IBR's LVRT and recovery on rotor dynamics of nearby SGs, the cases in TABLE II. are discussed. As shown in Fig. 5, during LVRT, IBR's grid-side converter (GSC) generates reactive current based on voltage droop control[6], while the active output is limited. In the recovery stage, reactive support is withdrawn immediately, and active current is restored using a ramp-up strategy. Different active current and recovery rates are discussed in TABLE II.

TABLE I
SYSTEM PARAMETERS

| Component | Item (Symbol) | Value |
| --- | --- | --- |
| GSC | Proportional and integral gain | PLL | 50,900 |
| | | Inner loop controller | 1,10 |
| | | Outer loop controller | 0.5,10 |
| SG | Capacity | 600MVA |
| | Transient reactance ($X_d'$, $X_q'$) | 0.15,0.25 |
| | $D$, $T_J$ | 10,8s |
| Admittance | $Y_s$, $Y_w$, $Y_g$ | -j2.18, -j10, -j1.08 |
| Grid | Fundamental frequency ($f_g$) | 50Hz |
| | Base capacity | 1000MVA |

Case 1-4 are designed to validate the impacts of IBR's LVRT control on the rotor dynamics and transient stability of SG, with results presented in Fig. 8. In TABLE II, $\sigma$ represents the retention level of active current. As $\sigma$ decreases, $\Delta P_{wf}$ also decreases. Results of Cases 1 and 3 in Fig. 8(a) and (c) confirm the scenario described in Fig. 4(a): If $\Delta\omega$>0 during the fault, reducing $\Delta P_{wf}$ (or $\sigma$) introduces positive damping, enhancing SG's transient stability by lowering transient energy. Conversely, Case 2 and 4 (Fig. 8(b) and (d)) correspond to Fig. 4(b): If $\Delta\omega$<0 during the fault, a decrease in $\Delta P_{wf}$ (or $\sigma$) results in a negative damping effect, making the SG more prone to lose stability in the second swing.

TABLE II
STUDY CASES

| Case | IBR output (MW) | SG output (MW) | Transition resistance (Ω) | Active current reserve ($\sigma$) (p.u.) | Recovery rate (p.u./s) |
| --- | --- | --- | --- | --- | --- |
| 1 | 650 | 50 | 0 | 0.5,0.7 | ∞ |
| 2 | 650 | 50 | 0 | 0,0.1 | ∞ |
| 3 | 450 | 250 | 5 | 0.7,1.2 | ∞ |
| 4 | 450 | 250 | 25 | 0,0.5 | ∞ |
| 5 | 650 | 50 | 0 | 0.7 | ∞,5 |
| 6 | 650 | 50 | 0 | 0.5 | ∞,1 |
| 7 | 650 | 50 | 0 | 0.2 | ∞,0.4 |

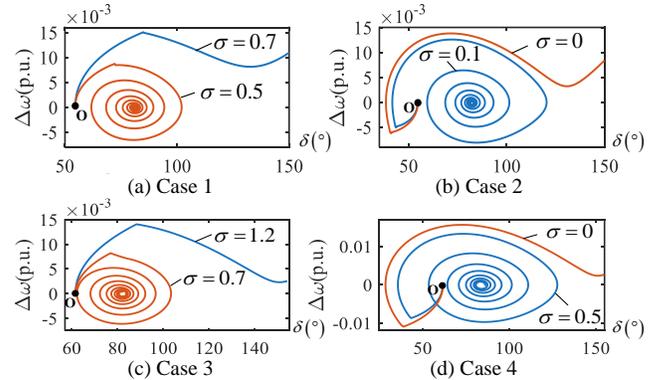

**Fig. 8.** Results of IBR's damping effect during LVRT.

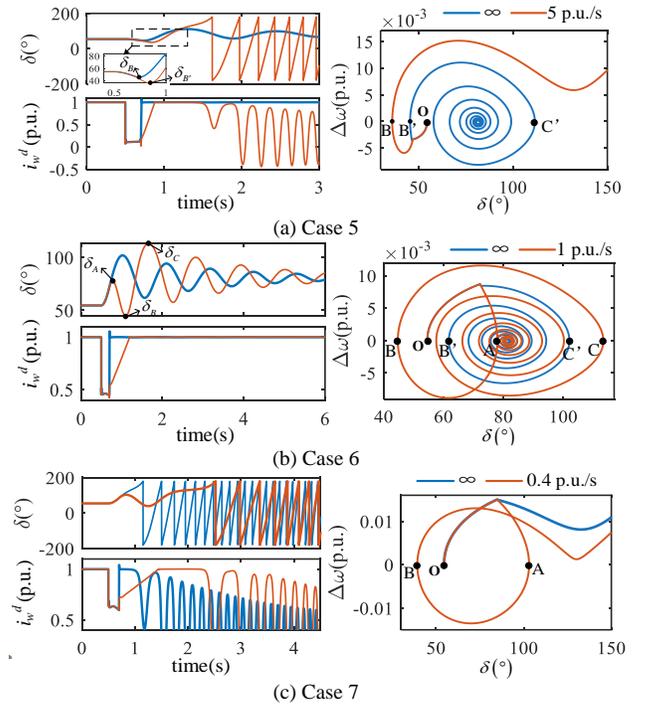

**Fig. 9.** Results of IBR's damping effect during LVRT recovery.

Cases 5-7 concentrate on examining how IBR's LVRT recovery control impacts the rotor dynamics of SG. Fig. 9 presents SG's rotor angle, phase trajectory, and IBR's active current. As depicted in Fig. 9(a) and (b), regardless of whether

SG accelerates or decelerates during faults, a slower restoration of $P_w$ in the angle-decreasing swings post-fault leads to a greater accumulation of transient energy compared to immediate recovery. This is evident from the reduced minimum rotor angle of SG (from B' to B), indicating a closer proximity to the stability boundary. In the subsequent swing with $\Delta\omega>0$, due to the nearly complete recovery of IBR's active current, the decrease of energy caused by IBR's positive damping effect in this swing is less than its increase in the last swing. Therefore, the transient energy increases at the end of one cycle, pushing the SG towards a potential multi-swing acceleration instability. Specifically, Case 7 (Fig. 9(c)) demonstrates that the SG's instability swing is delayed from the first swing to the third swing under the slow recovery control. These results validate the previous analysis.

## V. Discussion

The above analysis reveals the mechanism of IBR's LVRT and recovery control on the transient energy of SGs and verifies the existence of multi-swing instability through simulations. However, the manifestation of this phenomenon depends on various factors, including but not limited to the electrical distance between IBR and SG, the output of IBR and SG, the type of fault, and the physical damping of SG. Firstly, a higher proportion of IBR output and closer electrical proximity between the IBR and SG makes the rotor dynamics of SG more sensitive to changes in $P_w$ during the transients, thus increasing the likelihood of multi-swing instability. Additionally, different fault locations and types significantly affect SG dynamics during LVRT. In the presence of transition resistance, SGs are more likely to decelerate during a fault. Finally, the impact of SG's physical damping is pronounced during the LVRT recovery process. Increasing SG damping not only benefits its first-swing stability but also can improve the multi-swing stability if it offsets the negative damping effect of IBR.

## VI. Conclusion

This letter offers an analytical perspective on how GFL-IBRs' LVRT and recovery control affect the transient stability of nearby SGs. Notably, IBR introduces a time-varying, non-uniform damping torque, altering the unbalanced power during different rotor swing stages of nearby SGs and potentially leading to multi-swing instability. Mitigating these negative effects of IBRs and enhancing the overall stability of the system will be the focus of our future work.

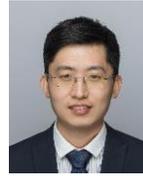 **Songhao Yang** (Senior Member, IEEE) was born in Shandong, China, in 1989. He received the B.S. and Ph.D. degrees in electrical engineering from Xi'an Jiaotong University, Xi'an, China, in 2012 and 2019, respec-tively, and the Ph.D. degree in electrical and electronic engineering from Tokushima University, Tokushima, Japan, in 2019. He is currently an Associate Professor with Xi'an Jiaotong University. His research focuses on power system stability analysis and control.

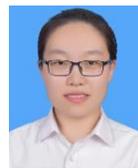 **Bingfang Li** (Student Member, IEEE),, received the B.S. degree from North China Electric Power University, Baoding, China, in 2022, and is currently working toward the Ph.D. degree with Xi'an Jiaotong University. Her main fields of interest include Power system stability analysis and control.

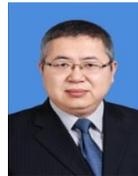 **Zhiguo Hao** (Senior Member, IEEE), was born in Ordos, China, in 1976. He received the B.Sc. and Ph.D. degrees in electrical engineering from Xi'an Jiaotong University, Xi'an, China, in 1998 and 2007, respectively. He is currently a Professor with the Electrical Engineering Department, Xi'an Jiaotong University. His research focuses on power system protection and control.

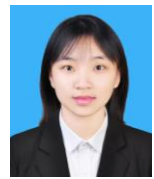 **Yiwen Hu**, received the B.S. degree from North China Electric Power University, Baoding, China, in 2023, and is currently working toward the M.S. degree with Xi'an Jiaotong University. Her main fields of interest include Power system stability analysis and control.

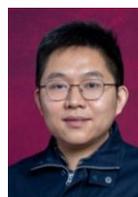 **Huan Xie**, is working as an electrical engineer at State Grid Jibei Electric Power Research Institute, Beijing, China. He received the B.Sc. and M.Sc. in Electrical Engineering and Automation from Hohai University in 2001 and 2014, and received Ph.D. degree in Electrical Engineering and Automation from Xi'an Jiao Tong University, Xi'an .China in 2008. His areas of interest include power system stability and control.

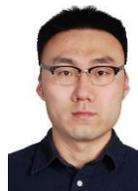 **Tianqi Zhao**, is working as an electrical engineer at State Grid Jibei Electric Power Research Institute, Beijing, China. He received the B.Sc., M.Sc. and Ph.D. degree in Electrical Engineering and Automation from Tianjin University, Tianjin .China, in 2012, 2014 and 2018 respectively. His areas of interest include power system stability and control, renewable energy integration and analytical methods for planning and operations.